\documentclass[10pt,conference]{IEEEtran}
\IEEEoverridecommandlockouts

\usepackage{cite}
\usepackage{amsmath,amssymb,amsfonts}
\usepackage{url}
\usepackage{algorithmic}
\usepackage{graphicx}
\usepackage{textcomp}
\usepackage{booktabs}
\usepackage{verbatim}
\usepackage{xcolor}
\usepackage{colortbl}
\definecolor{summgray}{gray}{0.92}
\def\BibTeX{{\rm B\kern-.05em{\sc i\kern-.025em b}\kern-.08em
    T\kern-.1667em\lower.7ex\hbox{E}\kern-.125emX}}

\makeatletter
\newcommand{\linebreakand}{%
  \end{@IEEEauthorhalign}
  \hfill\mbox{}\par
  \mbox{}\hfill\begin{@IEEEauthorhalign}
}
\newcommand{\spm}[1]{{\scriptsize$\,\pm\,$#1}}
\newcommand{\B}[1]{\textbf{#1}}
\makeatother

\begin{document}
\title{Cross-Layer Intrusion Detection in 5G O-RAN: Gains and Limits of Fusing Radio Telemetry with Network Flow Records\\
\thanks{HF and GW were supported by the Federal Ministry of Education and Research of Germany (BMBF) within “6G-RIC: 6G Research and Innovation Cluster”, under project identification number 16KISK025, and the BMBF joint project “UltraSec: Security Architecture for UWB-based Application Platform”, project identification number 16KIS1682. GW was supported by the German Science Foundation (DFG) within priority program SPP 2378: “ResNets: Resilience in Connected Worlds” under grant WU 598/12-1.}
}

\author{%
\IEEEauthorblockN{ Hamed Fard}
\IEEEauthorblockA{\textit{Department of Computer Science} \\
\textit{Freie Universität Berlin}\\
Berlin, Germany \\
h.habibi.fard@fu-berlin.de}
\and
\IEEEauthorblockN{Ilya Komarov}
\IEEEauthorblockA{\textit{Bundesdruckerei GmbH} \\
\textit{Abteilung Innovations}\\
Berlin, Germany \\
ilya.komarov@bdr.de}
\and
\IEEEauthorblockN{Gerhard Wunder }
\IEEEauthorblockA{\textit{Department of Computer Science} \\
\textit{Freie Universität Berlin}\\
Berlin, Germany \\
g.wunder@fu-berlin.de}
%
%

}

\maketitle


\begingroup
\renewcommand\thefootnote{}
\footnotetext{\textcopyright~2026 IEEE. Personal use of this material is permitted. Permission from IEEE must be obtained for all other uses, in any current or future media, including reprinting/republishing this material for advertising or promotional purposes, creating new collective works, for resale or redistribution to servers or lists, or reuse of any copyrighted component of this work in other works.}
\endgroup

\begin{abstract}
Open RAN disaggregation enables joint analysis of DU radio telemetry and CU-side network-flow records, motivating cross-layer intrusion detection. We evaluate whether fusing these two modalities improves over each individually across seven architectures, using run-disjoint splits over ten seeds on a live 5G O-RAN dataset. Radio features match or outperform network flows on ROC-AUC and run-level detection
rate across all architectures. Fusion yields selective ROC-AUC gains but at a one-percent false-positive operating point improves detection rate only for GRU and Transformer, reducing it for the other five models. The benefit is confined to architectures where both single-modality detection rates fall below 0.75. A DoS-to-Benign confusion of 27 to 46 percent persists across all 42 tested configurations of architecture, modality, and window duration, pointing to a limitation in the tested windowed statistical aggregation rather than in model capacity. Code is publicly available.\footnote{\url{https://github.com/afbf4c8996f/cross-layer-ids-5g-oran}}
\end{abstract}


\section{Introduction}

Open Radio Access Network (O-RAN) architectures split traditional base-station functions across CU, DU, and RU components connected by open interfaces~\cite{polese2023understanding}.
A programmable Near-RT RIC hosts third-party xApps for closed-loop network control, and recent extensions allow distributed applications (dApps) to run directly at the DU~\cite{abdalla2024end}. This flexibility also means that intrusion detection can draw on telemetry from multiple collection points. CU-side monitoring provides network-flow information on transport and application behavior, while DU-side monitoring exposes radio-layer indicators such as signal quality, modulation, and block error rates.
These two sources differ in sampling rate, timing, and coverage, and they are not guaranteed to be aligned at the packet level.

The central question for O-RAN intrusion detection is therefore practical: when DU radio telemetry already captures attack signatures, does adding CU flow evidence improve detection, and under what conditions does combining the two help or hurt?
Answering this requires training models on each modality separately, fusing their outputs, and evaluating across multiple attack families and repeated run-disjoint partitions to separate consistent gains from partition-specific effects.
We therefore adopt a windowed representation in which each modality is aggregated into fixed-duration statistical summaries, a standard design choice in time-aware IDS~\cite{parlanti2025temporal} that yields fixed-size inputs with a latency-context trade-off suited to O-RAN monitoring.

Prior work on O-RAN security has addressed protocol-specific threats, radio-side monitoring, and cross-domain label transfer (Section~\ref{sec:related}).
A direct comparison of standalone CU features, standalone DU features, and their score-level fusion on a shared paired dataset has not been systematically evaluated.

This paper addresses that gap using the NetsLab-5GORAN-IDD dataset~\cite{zadeh2025descriptor}, which pairs CU Zeek flow records with DU radio telemetry across 42 experiment runs covering Benign and five attack families. Each modality is preprocessed independently and aggregated into overlapping time windows.
Seven architectures are trained per modality. For binary detection, two score-level fusion strategies are evaluated. All partitions are run-disjoint, meaning every window from a given run appears in exactly one of train, validation, or test, and the procedure is repeated over 10 random seeds. Binary detection is reported as ROC-AUC and as a run-level detection rate at a 1\% false-positive operating point. Multiclass classification is reported as F1-macro over the six traffic types.

The main findings are as follows.
Radio telemetry matches or exceeds network flows for binary detection across all seven architectures.
Fusion can preserve or slightly improve ROC-AUC, but at a fixed low false-positive threshold it can reduce run-level detection rate when the weaker modality dilutes borderline scores.
For multiclass classification, mean-probability fusion improves F1-macro for every model tested.
At the class level, fusion benefits some categories substantially (Web~$F_1$ improves by $+0.222$) while DoS remains a persistent bottleneck, with a DoS-to-Benign confusion that persists across architectures, feature sets, and window durations.

The contributions of this work are:
\begin{itemize}
  \item A cross-layer intrusion detection study on paired O-RAN CU
    flow records and DU radio telemetry, evaluated under a
    run-disjoint protocol over 10 seeds with seven model
    architectures.
  \item A score-level fusion analysis that jointly reports ROC-AUC, run-level detection rate at 1\% FPR, and time-to-detection, showing that threshold-free gains do not necessarily translate into better low-FPR operational detection.
  \item A per-class error analysis identifying a
    DoS-to-Benign confusion that persists across 42
    model-modality-window configurations.
\end{itemize}

Section~\ref{sec:related} reviews related work.
Section~\ref{sec:method} describes the dataset, preprocessing,
models, fusion, and evaluation protocol.
Section~\ref{sec:results} presents the results, and
Section~\ref{sec:conclusion} concludes.

\section{Related Work}
\label{sec:related}

A recent survey of intrusion detection in Open~RAN~\cite{amachaghi2024survey}
highlights that disaggregated components and programmable
interfaces create both new attack surfaces and new monitoring
opportunities.
Within O-RAN, prior work follows two main directions.

The first focuses on security services embedded in the O-RAN
control loop.
An O-RAN-compliant Layer-3 detection framework is proposed
in~\cite{wen20245g}, where a rule-based expert system
deployed as an xApp detects RRC and NAS protocol exploits
using fine-grained cellular telemetry.
A different approach appears in~\cite{scalingi2024det},
which targets real-time RRC attack detection through PHY and
cross-layer features at the network edge and evaluates on
scenarios not seen during training.
In~\cite{tsourdinis2024ai}, DoS detection is combined
with dynamic resource allocation through an AI-driven xApp
on a live O-RAN testbed.
These studies show that O-RAN can host practical detection
logic.
Their emphasis, however, is on protocol-specific exploits,
control-plane threats, or online mitigation of specific attacks
rather than on comparing how different telemetry sources
perform across multiple attack families.

The second direction is closer to the cross-layer question
examined here.
DU-side air-interface measurements are used
in~\cite{xavier2023machine} to detect denial-of-service traffic
before it reaches the CU, with classifiers operating on
per-second physical and MAC layer records.
Within the NetsLab-5GORAN-IDD
setting~\cite{zadeh2025descriptor},~\cite{civcisssilent}
compares dApp-based radio-side detection at the DU
with xApp-based packet inspection at the CU on the same testbed, but trains and evaluates each modality independently without temporal alignment, paired scoring, or fusion, and reports a single train-test split without run-disjoint partitioning or multi-seed significance analysis.

A cross-domain approach in~\cite{xavier2024cross}
transfers labels from a transport-network classifier to train a RAN-side KPI classifier, but does not evaluate each domain
independently or combine their output scores.
Outside O-RAN, multimodal IDS methods combine
heterogeneous input representations to improve
detection~\cite{shi2023multimodal,he2019novel}.
Their modalities, however, are typically multiple views of the
same network traffic rather than CU and DU measurements
collected from different O-RAN components.

Taken together, prior work establishes that radio-side
measurements carry useful attack information and that O-RAN
can support deployable detection services.
What remains less explored is a direct comparison of CU
network-flow features and DU radio telemetry on the same
paired dataset, including the conditions under which fusion
helps and where one modality alone is already sufficient. Section~\ref{sec:method} describes the dataset, evaluation
protocol, and fusion strategy used to address these gaps.

\section{Methodology}
\label{sec:method}

\subsection{Dataset, Feature Spaces, and Alignment}
\label{subsec:dataset}

We evaluate on the NetsLab-5GORAN-IDD
dataset~\cite{zadeh2025descriptor}, collected from a live
5G Open~RAN testbed at University College Dublin built on
the OpenAirInterface platform.
Each experiment run produces two parallel data streams.
Network packets are captured at the Centralized Unit (CU)
and processed into Zeek flow records
(\texttt{conn.log}, \texttt{http.log}, \texttt{files.log}).
Radio telemetry is aggregated at one-second intervals across all UEs active in the DU. Only the aggregate UE and row count summaries are preserved. No individual UE identifiers are used as model inputs. The traffic includes six families (Benign, DoS, DDoS, Probe, BruteForce, Web). For study, we include 42~paired runs for which both modalities are available.

\textbf{Network features.}
Zeek flow records contain per-flow duration, byte and packet
counts in each direction, transport protocol, application-layer
service, TCP connection state, TCP flag history, and
HTTP-level indicators. Windowing and preprocessing
(Section~\ref{subsec:preprocessing}) map these fields to
68 to 72~features per window across seeds after feature ablation.
The exact dimension varies slightly because one-hot vocabularies are fit on the training partition for each seed.

\textbf{Radio features.}
The lower-layer telemetry includes signal quality measurements
(RSRP, RSRQ, RSSI, SINR), modulation and coding parameters
(MCS, CQI), block error rates, transmission power, SNR,
and byte counters.
After windowing and preprocessing, the radio side has
44~features per window.
Among these are missingness indicators that record the
fraction of seconds without telemetry. If a KPI aggregate is undefined due to missing telemetry, it is mapped to a fixed finite value in the model inputs, while the missingness indicators preserve whether the value came from an observed or missing window.

\textbf{Alignment.}
The two modalities come from different O-RAN components and
are sampled differently.
We do not assume exact timestamp agreement between them.
Instead, all timestamps within a run are expressed relative
to the start of that run.
The network windowing step defines the window grid, and the
radio step reuses it directly.
A network window at $[t, t{+}W)$ and the corresponding radio
window therefore cover the same interval on the run timeline.
Some DoS traces lack telemetry timestamps entirely.
For these, time is assigned from the sample index at one hertz,
which preserves ordering without requiring absolute clocks.
Manual inspection of the affected traces confirmed that the
one-hertz assumption is consistent with the nominal telemetry
sampling rate of the remaining runs.
Windows where the radio stream has gaps are not discarded
but annotated with the missingness indicators described above.
Score-level fusion (Section~\ref{subsec:fusion}) avoids requiring per-packet synchronization between modalities.

\subsection{Preprocessing and Windowing}
\label{subsec:preprocessing}

Sliding windows of $W$ seconds with stride $S$ convert
variable-length runs into fixed-size samples
($W \in \{5, 10\}$, $S = 2$), producing overlapping windows.

Flows are assigned to windows by their start timestamp, and windows with no Zeek flows are discarded during preprocessing. A window is labeled Attack if at least half of its flows are labeled Attack.
This threshold assigns the label of the majority flow type in each window, avoiding sensitivity to a small number of misaligned flows at window boundaries. For multiclass, benign windows take label Benign, and attack windows take the modal attack family among the attack labeled flows in that window. The two modalities require different preprocessing.
On the network side, numeric flow fields are aggregated per
window as sum, mean, and standard deviation, categorical fields
become one-hot encoded modes, and heavily skewed features are
log-transformed before standardisation.
On the radio side, byte counters are cumulative and first
differenced into per-second rates before aggregation.
The remaining KPIs are summarised as mean and standard deviation
per window. All transforms are fit on the training partition only.
Full preprocessing details are available in the 
repository.

\subsection{Run-Disjoint Evaluation Protocol}
\label{subsec:run_disjoint}

Row-level or stratified random splitting allows windows from the
same run to appear in both training and test sets.
On this dataset, that causes 41 of 42 runs to leak across
partitions, letting the model partly match run-level conditions
rather than learn attack patterns that transfer to new traffic.
To prevent this, all windows from a given run are placed in
exactly one partition.
The 42~runs are split into train, validation, and test at
roughly 70/15/15 by run count, stratified by attack family
with at least one run per family in each partition. Across 10 seeds, the split yields 19 to 26 training runs, 7 to 12 validation runs, and 8 to 12 test runs, with at least two runs per family in training and at least one run per family in validation and test. This procedure is repeated with 10 random seeds to quantify sensitivity to partition composition.

\subsection{Models}
\label{subsec:models}

This work evaluates a heterogeneous set of seven architectures.
Three are classical machine learning models: logistic regression
(LogReg), gradient-boosted trees (XGBoost)~\cite{chen2016xgboost}, and random forest (RF).
Three are sequential deep learning models: a gated recurrent unit
(GRU)~\cite{cho2014learning}, a temporal convolutional network (TCN)~\cite{bai2018empirical}, and a Transformer~\cite{vaswani2017attention}.
The remaining model is a residual multi-layer perceptron (ResMLP)~\cite{he2016deep},
a non-sequential deep learning architecture.
For sequential models, each sample for window $i$ consists of the $L$ most recent windows ending at $i$ within the same run, and the target is the label of window $i$.
If fewer than $L$ past windows exist, the sequence is left padded and a padding indicator feature marks padded steps.
Hyperparameters for all seven models, including the sequence
length $L{=}8$ for sequential architectures, are tuned with
Optuna~\cite{akiba2019optuna} (50 trials, TPE sampler, median
pruning) on a single reference split (seed~43, $W{=}10$) and
then held fixed across all 10 evaluation seeds and both window
sizes. Tuning on a single seed reduces computational cost but
does not guarantee that the selected configuration is optimal
for every partition. Per-seed tuning would require 560
independent Optuna sweeps and risk overfitting hyperparameters
to individual partitions. The 10-seed evaluation quantifies
the resulting variance. The frozen configuration nonetheless
transfers to $W{=}5$ without re-tuning, with 4 of 7 models
achieving higher F1-macro at $W{=}5$ than at the tuned
$W{=}10$ (Section~\ref{subsec:multiclass}). Keeping
hyperparameters fixed across seeds and window sizes makes
the $W{=}5$ to $W{=}10$ comparison reflect window duration
and temporal aggregation rather than re-tuning. Each model
is trained independently per seed, per modality, and per task.
Deep learning models use early stopping on the validation set,
while classical models train to completion. The full set of
tuned hyperparameters is available in the companion repository.

\subsection{Fusion}
\label{subsec:fusion}

Network and radio models are trained separately and each
produces its own probability estimate.
Fusion operates on these scores, not on raw features.
For binary detection, two strategies are compared.
The first is a simple arithmetic mean of the two base
probabilities. The second trains a logistic regression meta-classifier
on the outputs of both base models. To avoid information leakage, the meta-classifier is not trained on in-sample predictions. Instead, out-of-fold predictions from the training partition are used. The meta-classifier therefore never sees outputs that the base models were fitted on. The out-of-fold partitions are constructed at the run level, placing all windows from a given run in exactly one fold. Multiclass fusion uses mean probability only.

Feature-level fusion is not evaluated because the two
modalities lack packet-level alignment and differ in
sampling rate, making concatenation of raw feature vectors
dependent on interpolation assumptions that would confound
the modality comparison. The two score-level strategies
tested here isolate the question of whether combining
output probabilities adds value over single-modality
predictions, without introducing architectural coupling
between the base models.

\subsection{Evaluation Metrics}
\label{subsec:metrics}

Binary detection is measured by ROC-AUC averaged over
10 seeds. At the operational level, a threshold is selected from training out of fold scores to achieve 1\% window level FPR on benign windows from benign runs. The same threshold is then applied unchanged to the held out test partition. An attack run counts as detected if at least one of its
windows exceeds this threshold, and the detection rate (DR) is the fraction of attack runs detected. An analogous benign-run false alarm rate is not reported because the dataset contains 10 benign runs out of 42, and stratified splitting at 70/15/15 allocates only one to the test partition per seed. A run-level false alarm estimate from a single benign run is binary (0\,\% or 100\,\%) and not meaningful. 
Multiclass classification is evaluated by F1-macro, the unweighted mean of per-class F1 across all six traffic types.

\section{Results}
\label{sec:results}

\subsection{Binary Attack Detection}
\label{subsec:binary}

\begin{table*}[t]
\centering
\caption{Binary attack detection: ROC-AUC (mean\,$\pm$\,std, 10 seeds)
and detection rate at 1\,\% FPR (benign-only threshold).
\textbf{Bold} = best per model.  $W{=}10$, held-out test, stratified run-disjoint splits.}
\label{tab:binary_detection}
\footnotesize
\setlength{\tabcolsep}{4pt}
\renewcommand{\arraystretch}{1.15}
\begin{tabular}{@{}l
  c@{\hskip 5pt}c@{\hskip 5pt}c@{\hskip 5pt}c
  @{\hskip 12pt}
  c@{\hskip 5pt}c@{\hskip 5pt}c@{\hskip 5pt}c@{}}
\toprule
& \multicolumn{4}{c}{\textbf{ROC-AUC}\,($\uparrow$)}
& \multicolumn{4}{c}{\textbf{Detection Rate}\,($\uparrow$)} \\
\cmidrule(lr){2-5} \cmidrule(l){6-9}
\textbf{Model}
  & Net & Radio & Fus.\textsubscript{S} & Fus.\textsubscript{M}
  & Net & Radio & Fus.\textsubscript{S} & Fus.\textsubscript{M} \\
\midrule
LogReg
  & .844\spm{.056} & .961\spm{.028}
  & \B{.973}\spm{.023} & .948\spm{.052}
  & .848 & \B{.961} & .679 & .846 \\
XGBoost
  & .837\spm{.063} & .949\spm{.044}
  & \B{.960}\spm{.036} & .942\spm{.050}
  & .624 & \B{.932} & .500 & .689 \\
RF
  & .840\spm{.067} & .932\spm{.075}
  & .941\spm{.075} & \B{.944}\spm{.062}
  & .636 & \B{.916} & .600 & .659 \\
\midrule
ResMLP
  & .830\spm{.047} & \B{.944}\spm{.040}
  & .926\spm{.078} & .932\spm{.051}
  & .436 & \B{.830} & .371 & .748 \\
GRU
  & .851\spm{.050} & .936\spm{.039}
  & \B{.943}\spm{.037} & .938\spm{.033}
  & .633 & .746 & .598 & \B{.771} \\
TCN
  & .833\spm{.066} & \B{.945}\spm{.042}
  & .865\spm{.171} & .927\spm{.056}
  & .714 & \B{1.00} & .759 & .755 \\
Transformer
  & .827\spm{.059} & .925\spm{.051}
  & .834\spm{.182} & \B{.931}\spm{.040}
  & .671 & .671 & .632 & \B{.771} \\
\bottomrule
\end{tabular}
\\[3pt]
{\scriptsize
Fus.\textsubscript{S}\,=\,stacked meta-learner;\;
Fus.\textsubscript{M}\,=\,probability averaging.\;
Detection rate\,=\,fraction of attack runs with $\geq$\!1 detected window.\;
Horizontal rule separates classical ML (top) from deep learning (bottom).}
\end{table*}

Table~\ref{tab:binary_detection} reports ROC-AUC and Detection Rate (DR) for seven model architectures across four feature configurations, namely network-only (Net), radio-only (Radio), stacked fusion (Fus.\textsubscript{S}), and mean fusion (Fus.\textsubscript{M}). All results use $W{=}10$ and are averaged over 10 run-disjoint seeds.

\textbf{Radio advantage.} Across all architectures, radio features match or exceed network flows on both metrics. The ROC-AUC gap ranges from $+8.5$ to $+11.7$ points and is model-agnostic. All seven architectures reach a narrow network range ($0.827$--$0.851$), suggesting an information bottleneck in the network features rather than a modeling limitation. Radio features reach $0.925$--$0.961$.

\textbf{Fusion trade-off.} For discrimination (ROC-AUC), Fus.\textsubscript{S} achieves the highest overall AUC among classical models (LogReg, $0.973$), while Fus.\textsubscript{M} exceeds radio-only AUC for 3/7 models. However, at the selected 1\% window-level FPR operating point (threshold selected from training out-of-fold
scores using benign windows from benign runs), fusion degrades performance. Fus.\textsubscript{M} lowers DR for 5/7 models, with losses of $8$--$26$ percentage points relative to radio-only. Fus.\textsubscript{S} is particularly unstable for TCN and Transformer, with ROC-AUC standard deviations of $\pm 0.171$ and $\pm 0.182$ across
seeds, and corresponding DR below their radio-only baselines. The meta-learner is trained on out-of-fold scores from
K-fold base models and evaluated on scores from base models
retrained on the full training partition. At 42 runs, a
difference of one or two runs between seeds shifts the base
model output distributions enough to alter the meta-learner
decision boundary.

This gap between ROC-AUC and run-level DR is consistent with late-fusion dilution at a fixed low-FPR operating point. Averaging a strong modality with a weaker one can reduce the margin of borderline attack windows, increasing missed detections. Fus.\textsubscript{M} improves DR only for the two models where radio-only DR is already the lowest in the table (GRU $0.746$, Transformer $0.671$), suggesting that in this benchmark, fusion benefits detection primarily when the radio signal alone is insufficient. The Transformer yields the same seed-mean DR for Net and Radio in Table~\ref{tab:binary_detection} ($0.671$). Radio KPIs are natively available at the near-RT RIC via the E2 interface. Radio-only IDS deployment therefore requires no
additional DPI or flow monitoring infrastructure. 

\begin{figure*}[tp]
\centering
\includegraphics[width=\textwidth]{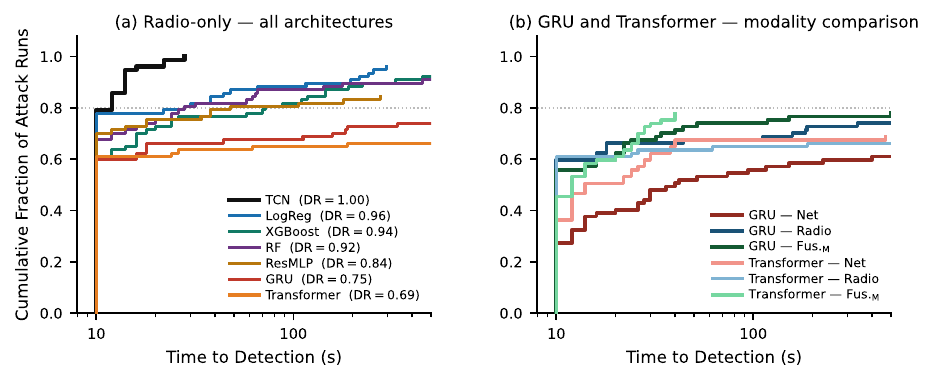}
\caption{Time-to-detection CDFs at 1\,\% FPR (benign-calibrated threshold,
  flow-onset, $W{=}10$, 10 run-disjoint seeds pooled).
  Each curve aggregates attack runs pooled across 10 seeds.
  The y-axis termination of each curve is the detection rate over all attack runs pooled across 10 seeds and differs from the seed-mean DR in Table~\ref{tab:binary_detection} because a CDF requires individual run observations, not seed-level aggregates.
  (a)~Radio-only features across all seven architectures.
  TCN reaches DR\,=\,1.00 with 79\,\% of attack runs detected within the first window (10\,s). The three classical and one non-sequential deep-learning model
  (LogReg, XGBoost, RF, ResMLP) cluster between
  DR\,=\,0.84 and DR\,=\,0.96. GRU and Transformer (red and orange) terminate below the 0.80 reference and are examined further in panel~(b).
  (b)~GRU and Transformer under three feature configurations.
  Fus.\textsubscript{M} raises DR above both single-modality baselines for both models, matching the two cases in
  Table~\ref{tab:binary_detection} where fusion improves detection rate.}
\label{fig:ttd_cdf}
\end{figure*}

\subsection{Detection Latency}
\label{subsec:ttd}

Fig.~\ref{fig:ttd_cdf} reports the time-to-detection (TTD)
distribution for all configurations, measured from first attack
flow to the first window exceeding the benign-calibrated threshold at 1\,\% FPR. Each curve aggregates attack runs pooled across 10 seeds. The y-axis termination of each curve is the detection rate over all attack runs pooled across 10 seeds and differs from the seed-mean DR in Table~\ref{tab:binary_detection} because a CDF requires individual run observations, not seed-level aggregates.

Panel~(a) shows radio-only performance across all seven
architectures. TCN detects 79.2\,\% of attack runs within the
first 10-second window and reaches DR\,=\,1.00 by 60\,s. The
remaining four models between DR\,=\,0.84 and DR\,=\,0.96 follow a comparable temporal profile: 60\,\%--78\,\% of their detections fall at 10\,s. Detection latency is therefore determined primarily by whether the threshold is crossed at all, not by how quickly it is crossed once the signal is present.

Panel~(b) shows GRU and Transformer under all three feature
configurations. Network-only features produce the lowest DR and the slowest accumulation for both models. Fus.\textsubscript{M} raises DR to 0.779 for both GRU (from 0.753 radio-only) and Transformer (from 0.688 radio-only). In the pooled attack-run analysis of Fig.~\ref{fig:ttd_cdf}(b), Fus.\textsubscript{M} reaches the same endpoint DR (0.779) for both GRU and Transformer. The Fus.\textsubscript{M} curves reach their termination points within the same latency range as the corresponding radio-only curves, with Transformer Fus.\textsubscript{M} completing all detections by 60\,s.

\begin{table}[t]
\centering
\caption{Multiclass traffic-type classification (F1-macro, mean\,$\pm$\,std,
10 seeds).  Six classes: Benign, BruteForce, DDoS, DoS, Probe, Web.
$\Delta$\,=\,Fus.\textsubscript{M}\,$-$\,Radio.
\textbf{Bold}\,=\,best per model.
$W{=}10$, held-out test, stratified run-disjoint splits.}
\label{tab:multiclass}
\footnotesize
\setlength{\tabcolsep}{3pt}
\renewcommand{\arraystretch}{1.15}
\begin{tabular}{@{}l ccc r@{}}
\toprule
\textbf{Model}
  & Net & Radio & Fus.\textsubscript{M}
  & \multicolumn{1}{c}{$\Delta$} \\
\midrule
LogReg
  & .380\spm{.119} & .471\spm{.086}
  & \B{.579}\spm{.123} & \emph{+.108} \\
XGBoost
  & .389\spm{.114} & .619\spm{.122}
  & \B{.683}\spm{.116} & \emph{+.064} \\
RF
  & .363\spm{.108} & .609\spm{.122}
  & \B{.670}\spm{.076} & \emph{+.061} \\
\midrule
ResMLP
  & .361\spm{.097} & .518\spm{.095}
  & \B{.587}\spm{.120} & \emph{+.068} \\
GRU
  & .388\spm{.096} & .521\spm{.122}
  & \B{.604}\spm{.118} & \emph{+.083} \\
TCN
  & .364\spm{.095} & .486\spm{.104}
  & \B{.530}\spm{.106} & \emph{+.043} \\
Transformer
  & .391\spm{.092} & .517\spm{.104}
  & \B{.593}\spm{.106} & \emph{+.076} \\
\bottomrule
\end{tabular}
\\[3pt]
{\scriptsize
Fus.\textsubscript{M}\,=\,probability averaging.\;
$\Delta$\,=\,absolute gain from fusion (all positive).\;
Horizontal rule separates classical ML from deep learning.}
\end{table}

\subsection{Multiclass Traffic-Type Classification}
\label{subsec:multiclass}

Table~\ref{tab:multiclass} reports F1-macro for the six-class traffic-type classification task (Benign, BruteForce, DDoS, DoS, Probe, Web). Only mean-probability fusion (Fus.\textsubscript{M}) is evaluated here, as the multiclass pipeline uses direct train-to-predict without the out-of-fold predictions that stacked fusion requires.

Two patterns are consistent across all models in Table~\ref{tab:multiclass}. Radio features outperform network features by $+9.1$ to $+24.6$ F1-macro points, and Fus.\textsubscript{M} improves over the better single-modality result by $+4.3$ to $+10.8$ points ($\Delta$ column). XGBoost with Fus.\textsubscript{M} achieves the highest F1-macro at $0.683 \pm 0.116$. At $W{=}5$, F1-macro is higher for 4 of 7 models under Fus.\textsubscript{M}, with XGBoost reaching $0.724$ versus $0.683$ at $W{=}10$. All results above use $W{=}10$ to match the HPO
configuration.

The moderate absolute F1-macro should be interpreted relative to the evaluation protocol. Splits are run-disjoint, meaning every test run is withheld entirely from training. The model must therefore generalize to unseen run-level realizations
of each attack family, with no overlap between training and test runs. Classification also targets six classes at window level, where a uniform six-class chance baseline yields $F_1\text{-macro} = 1/6 \approx 0.167$ under balanced classes. 

Within-family heterogeneity in the dataset accounts for much of the variability across seeds. Across DoS runs appearing in the held-out folds, the number of DoS windows per run ranges from 4 to 838 (median 40). Depending on which runs fall into the test partition, DoS constitutes 10.7\% to 30.4\% of test windows across seeds, while Benign ranges from 13.2\% to 29.2\%. These composition shifts likely contribute to the observed F1-macro standard deviations of $0.076$--$0.123$.

\begin{figure*}[t]
\vspace{-1pt}
\centering
\includegraphics[width=\textwidth]{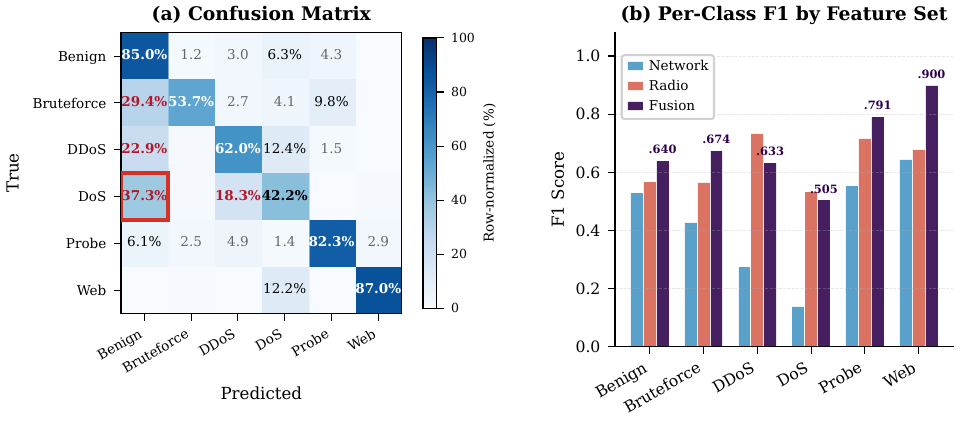}
\caption{Per-class analysis of XGBoost fusion\_mean
  ($W{=}10$, 10 run-disjoint seeds aggregated).
  (a)~Row-normalised confusion matrix. The red border highlights
  the dominant error: 37.3\% of DoS windows are classified as Benign.
  (b)~Per-class F1 by feature set. Per-class F1 values are computed from predictions pooled across all
10 seeds to stabilize rare-class cell counts in the confusion matrix. The F1-macro in Table~\ref{tab:multiclass} is the mean of per-seed macro-F1 scores and differs from the unweighted mean of the six values shown here.}
\label{fig:perclass}
\end{figure*}

\subsection{Per-Class Analysis}
\label{subsec:perclass}

Since XGBoost with Fus.\textsubscript{M} yields the highest multiclass F1-macro (Table~\ref{tab:multiclass}), it is used for the per-class analysis in Fig.~\ref{fig:perclass}.

Panel~(a) shows the row-normalised confusion matrix. The top six
confusion pairs account for 74.0\% of all misclassifications.
Attacks misclassified as Benign represent the dominant error
pattern at 52.9\% of all errors. The largest attack-to-Benign
confusions are DoS$\,\to\,$Benign (37.3\% of DoS windows),
BruteForce$\,\to\,$Benign (29.4\% of BruteForce windows), and
DDoS$\,\to\,$Benign (22.9\% of DDoS windows). Additional prominent
confusions are DoS$\,\to\,$DDoS (18.3\% of DoS windows) and
DDoS$\,\to\,$DoS (12.4\% of DDoS windows). The attack-to-Benign errors
are consistent with the difficulty of separating low-rate
or early-phase attack windows from benign traffic using
aggregated statistics at 10-second granularity. The DoS-DDoS confusion persists despite the network features including a window-level unique source-IP count
(\texttt{n\_unique\_src\_ip}), suggesting that this
aggregate measure alone does not reliably separate single-source from distributed floods at 10-second granularity.

Panel~(b) reports per-class F1 for each feature set. Fusion improves four of six classes. Web benefits the most ($+0.222$, from $0.678$ radio to $0.900$ fusion), as both network flows and radio KPIs capture complementary aspects of HTTP-based attacks. Probe ($+0.074$) and BruteForce ($+0.109$) also gain from fusion. For DoS and DDoS, however, fusion reduces per-class F1. DDoS drops from $0.734$ (radio) to $0.633$ (fusion), and DoS from $0.532$ to $0.505$. For XGBoost, network features contribute almost no discriminative signal for these two classes ($F_1$ of $0.139$ and $0.274$, respectively), so fusing them with radio adds noise rather than information. DoS is the overall bottleneck class at $F_1 = 0.505$, with 37.3\% of its windows misclassified as Benign.

This per-class pattern is consistent with the binary detection results in Table~\ref{tab:binary_detection}. DoS, BruteForce, and DDoS are the three classes most frequently misclassified as Benign, and such benign-like attack windows likely contribute to the run-level detection-rate degradation observed under fusion in Section~\ref{subsec:binary}.

\begin{table}[t]
\centering
\caption{DoS$\,\to\,$Benign confusion rate (\%). Each cell aggregates 10 run-disjoint seeds. The DoS$\to$Benign confusion remains substantial across all 42 configurations (27--46\%, mean\,=\,37\%), indicating a persistent limitation under this windowed statistical aggregation.}
\label{tab:invariance}
\vspace{2pt}
\footnotesize
\setlength{\tabcolsep}{3.6pt}
\begin{tabular}{@{} l ccc ccc @{}}
\toprule
 & \multicolumn{3}{c}{$W{=}5$ (5\,s)} & \multicolumn{3}{c}{$W{=}10$ (10\,s)} \\
\cmidrule(lr){2-4} \cmidrule(lr){5-7}
Model & Net & Radio & Fus.\textsubscript{M} & Net & Radio & Fus.\textsubscript{M} \\
\midrule
    RF          & 42.3 & 39.9 & 38.8 & 38.8 & 39.3 & 38.8 \\
    LogReg      & 46.3 & 39.7 & 41.8 & 40.6 & 36.7 & 39.8 \\
    XGBoost     & 41.5 & 39.6 & 37.9 & 37.6 & 37.5 & 37.3 \\
    \midrule
    GRU         & 27.3 & 37.9 & 32.6 & 32.4 & 36.1 & 36.3 \\
    ResMLP      & 37.4 & 36.9 & 39.2 & 35.1 & 36.3 & 36.9 \\
    TCN         & 27.1 & 36.4 & 35.1 & 30.8 & 36.0 & 34.3 \\
    Transformer & 28.3 & 36.2 & 35.2 & 33.5 & 35.3 & 34.4 \\
\midrule[\heavyrulewidth]
\rowcolor{summgray}
    Range       & 27--46 & 36--40 & 33--42 & 31--41 & 35--39 & 34--40 \\
\bottomrule
\end{tabular}
\vspace{1pt}

{\scriptsize\textit{Note:} Radio columns span only 4.6\,pp (35.3--39.9\%),
confirming the confusion is dominated by the feature representation,
not the classifier. Full per-class metrics and reproduction scripts
are available in the companion repository.}
\end{table}

\begin{figure*}[t]
\centering
\includegraphics[width=\textwidth]{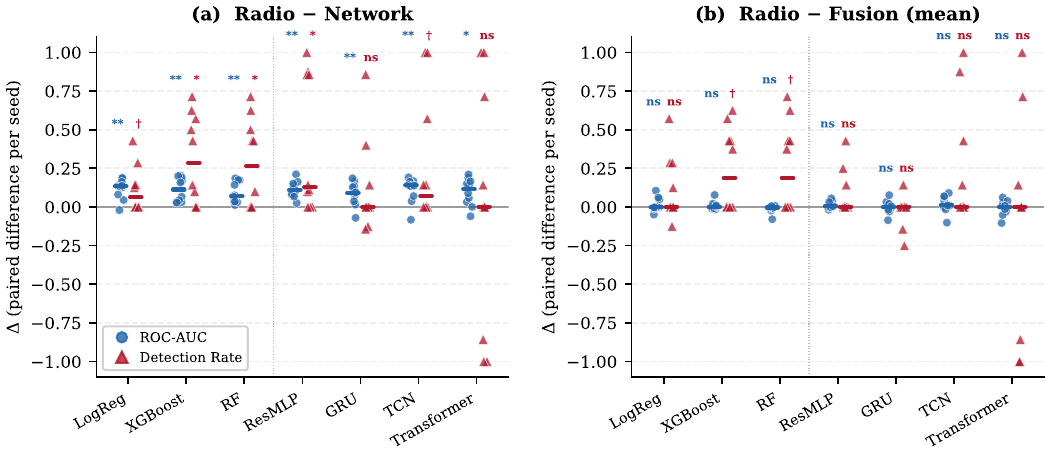}
\caption{Paired within-seed differences (Wilcoxon signed-rank test,
  two-sided, 10 run-disjoint seeds).
  Each dot is one seed's difference; circles (blue) show ROC-AUC,
  triangles (red) show detection rate at 1\,\% FPR.
  Horizontal bars indicate medians.
  Colored significance labels report $p$-values per metric
  (\textsuperscript{\textdagger}$p{<}0.10$,
   ${}^{*}p{<}0.05$,
   ${}^{**}p{<}0.01$).
  The dotted vertical line separates classical ML (left)
  from deep learning (right).
  (a)~Radio minus Network.
  ROC-AUC differences are significantly positive for all seven models
  (all $p{<}0.014$), confirming the radio advantage.
  DR differences are significant for XGBoost, RF, and ResMLP
  ($p{<}0.032$) and marginal for LogReg and TCN.
  (b)~Radio minus Fusion (mean).
  Neither metric shows a significant difference for any model,
  indicating that the DR changes reported in Table~\ref{tab:binary_detection}
  are directional but not statistically reliable at $n{=}10$.}
\label{fig:significance}
\end{figure*}

\subsection{Analysis of the DoS Classification Boundary}
\label{subsec:dos_boundary}

Table~\ref{tab:invariance} reports the DoS$\,\to\,$Benign confusion rate across 42 configurations (7 model architectures $\times$ 3 feature sets $\times$ 2 window sizes) to assess whether this confusion is tied to specific modeling choices or to the feature representation.

At $W{=}10$, the confusion rate ranges from 30.8\% to 40.6\% across all seven architectures, from logistic regression to Transformer. Since a linear classifier and an attention-based model produce comparable confusion rates, the bottleneck is unlikely to be model capacity. Across feature sets, both modalities yield nearly identical confusion for XGBoost (network 37.6\%, radio 37.5\%, fusion 37.3\%). When both modalities independently misclassify a comparable fraction of DoS windows as Benign, late fusion has little corrective signal to offer. The radio columns show the narrowest spread in the table at 35.3\%--39.9\% across all 14 radio configurations (7 models $\times$ 2 window sizes, std $\approx$ 1.5 pp), suggesting a consistent lower bound on radio-based separability for this class pair.

Reducing the window duration from 10\,s to 5\,s changes the confusion rate by only $0.6$ percentage points for XGBoost Fus.\textsubscript{M} (37.9\% vs 37.3\%), which suggests the limitation is not temporal resolution. Aggregated statistics computed over any fixed window discard the sequential ordering of flows within that window.

Across all 42 configurations, the DoS$\,\to\,$Benign confusion rate remains substantial (27--46\%, mean $\approx$ 37\%), indicating a persistent ambiguity in the feature representation that does not resolve with changes in architecture, modality, or window size.

\subsection{Fusion and Class-Level Performance}
\label{subsec:fusion_discussion}

Sections~\ref{subsec:perclass} and~\ref{subsec:dos_boundary} showed
that the DoS$\,\to\,$Benign confusion is consistent across modalities, architectures, and window sizes.
A low-rate DoS attack such as Slowloris, which maintains partial HTTP connections at minimum sending rate, generates per-window statistics that are difficult to distinguish from a lightly loaded benign period. For XGBoost, both modalities exhibit near-identical DoS$\,\to\,$Benign confusion rates (Net 37.6\%, Radio 37.5\%), and fusion remains essentially unchanged at 37.3\% (Table~\ref{tab:invariance}).

When both modalities assign a higher probability to Benign than to the true class for a given window, any nonnegative weighted combination of their outputs preserves that ordering. A non-monotonic meta-learner could in principle override its inputs, but Table~\ref{tab:invariance} shows substantial DoS$\,\to\,$Benign confusion for both modalities across architectures and window durations. The invariance across all tested configurations points to a persistent limitation in this windowed statistical aggregation rather than in model capacity.

Where the two modalities do encode different aspects of an attack, fusion is effective. Web attacks, for example, produce anomalies in both HTTP flow statistics and radio resource utilisation, and Fus.\textsubscript{M} improves Web $F_1$ by $+0.222$ over radio alone (Fig.~\ref{fig:perclass}).

\subsection{Statistical Reliability Across Seeds}
\label{subsec:significance}

To assess whether the observed differences are statistically
reliable across partition compositions, Fig.~\ref{fig:significance}
reports paired within-seed differences for ROC-AUC and detection
rate. Each comparison applies a two-sided Wilcoxon signed-rank
test~\cite{wilcoxon1945individual} over the 10 run-disjoint seeds.

Panel~(a) confirms that the radio advantage over network features
is significant for ROC-AUC across all seven architectures
($p{<}0.014$). Detection rate tells a more selective story.
The radio advantage reaches significance for XGBoost, RF, and
ResMLP ($p{<}0.032$), with LogReg and TCN marginal at
$p{=}0.063$. GRU and Transformer show no significant DR
difference between radio and network. For Transformer,
this is consistent with the identical seed-mean DR of
0.671 for both modalities in Table~\ref{tab:binary_detection}.
For GRU, the 11-point mean difference (0.746 vs 0.633)
does not reach significance due to high between-seed
variance.

Panel~(b) addresses whether fusion reliably changes performance.
It does not. ROC-AUC differences between radio and fusion are
uniformly non-significant (all $p{>}0.13$), and DR differences
do not reach significance for any model at the 0.05 level.
XGBoost and RF are marginal ($p{=}0.063$), reflecting the
directional DR loss visible in Table~\ref{tab:binary_detection} but
falling short of statistical confirmation with 10 partitions.

The high per-seed DR variance visible in both panels reflects
the sensitivity of a single-threshold operating point to
partition composition, consistent with the within-family
heterogeneity discussed in Section~\ref{subsec:multiclass}.
Taken together, these tests confirm that the radio advantage
is statistically robust for discrimination, while the fusion
effects on detection rate remain directional observations
that do not generalize beyond the observed partitions at
conventional significance levels.

\section{Conclusion}
\label{sec:conclusion}

Radio KPIs available at the near-RT RIC via the E2 interface
provide sufficient signal for binary intrusion detection across
all seven architectures evaluated. Score-level fusion yields
selective ROC-AUC gains but, at the selected 1\% window-level
FPR operating point, reduces detection rate for five of seven
models. Wilcoxon signed-rank tests confirm that the radio
advantage over network features is statistically significant
for ROC-AUC across all seven architectures ($p{<}0.014$),
while fusion-induced DR changes do not reach significance
at the 0.05 level. The benefit of mean-probability fusion is
confined to GRU and Transformer, the two architectures whose
radio-only detection rates fall below 0.75. For six-class
classification, fusion improves F1-macro for every model, with
the largest per-class gain on Web traffic. A DoS-to-Benign
confusion ranging from 27 to 46 percent persists across all
42 tested combinations of architecture, modality, and window
duration. Within the tested configurations, neither increasing
model capacity nor adding a second modality resolves this
confusion. Radio-only deployment requires no additional DPI
or flow monitoring infrastructure, which reduces integration
complexity at the near-RT RIC. To the best of our knowledge,
NetsLab-5GORAN-IDD is currently the only public O-RAN
dataset providing paired CU flow records and DU radio
telemetry. Confirming whether the patterns observed here
transfer beyond this testbed will require new paired
multi-layer captures from independent O-RAN deployments.

\bibliographystyle{IEEEtran}
\bibliography{references}

\end{document}